\documentclass[aps,nofootinbib,preprint]{revtex4}
\usepackage{graphicx}

\begin{document}
\def\dirac#1{#1\llap{/}}
\def\pv#1{\vec{#1}_\perp}

\title{B-Meson Wavefunction in the Wandzura-Wilczek Approximation}
\author{Tao, Huang$^{1,2}$\footnote{email:
huangtao@mail.ihep.ac.cn}, Xing-Gang, Wu$^{1}$\footnote{email:
wuxg@mail.ihep.ac.cn} and Ming-Zhen, Zhou$^{1}$\footnote{email:
zhoumz@mail.ihep.ac.cn}}
\address{$^1$CCAST(World
Laboratory), P.O.Box 8730, Beijing 100080, P.R.China,\\
$^2$Institute of High Energy Physics, Chinese Academy of Sciences,
P.O.Box 918(4), Beijing 100049, P.R. China.}

\begin{abstract}
The B-meson wavefunction has been studied with the help of the
vacuum-to-meson matrix element of the nonlocal operators in the
heavy quark effective theory. In order to obtain the
Wandzura-Wilczek-type B-meson wavefunction, we solve the equations
which are derived from the equation of motion of the light
spectator quark in the B meson by using two different assumptions.
Under the condition that $\omega_0=2\bar{\Lambda}$, the solutions
for the B-meson wavefunction in this paper agree well with the one
obtained by directly taking the heavy quark limit $m_b\to\infty$.
Our results show that the equation of motion of the light
spectator quark in the B meson can impose a strong constraint on
the B-meson wavefunctions $\Psi_\pm(\omega,z^2)$. Based on the
obtained results, we claim that both its distribution amplitudes
$\phi_B(\omega)$ and $\bar\phi_B(\omega)$ are important for
calculating the B meson decays. \\


\end{abstract}
\maketitle

\section{Introduction}

Along with the theoretical and experimental progresses, B physics
is attracting more and more attentions. The non-perturbative
light-cone (LC) wavefunction/distribtuion amplitude (DA) of the B
meson plays an important role in making reliable predictions for
exclusive B meson decays. The B meson DA has been investigated in
various
approaches\cite{bdistribution,braun,lange,qiao0,beneke,descotes,collinear}
and is the basis of the collinear factorization\cite{collinear}.
Ref.\cite{bdistribution} shows that the B meson DA is not
normalizable. Such feature of the B meson DA does not cause a
problem in practice\cite{lange}, but it does introduce an
ambiguity in defining the B meson decay constant $f_B$. Recently,
Ref.\cite{libwave} claims that it is the B meson wavefunction that
is more relevant to the B decays and in the framework of the
$k_T$-factorization theorem\cite{kt}, they proved that the B meson
wavefunction is renormalizable after taking into account
renormalization-group evolution effects. In the
$k_T$-factorization theorem, by taking into account the transverse
momentum dependence ($k_T$-dependence) into the non-perturbative
wavefunctions and the hard scattering, the endpoint singularity
(an example can be found in Ref.\cite{endpoint}) coming from the
collinear factorization can be cured. Theoretically, it is an
important issue to study the longitudinal and transverse momentum
dependence of the B wavefunction, since it provides a major source
of uncertainty in the calculations of the B decays.

Ref.\cite{qiao} presents an analytic solution for the B-meson
wavefunction, which satisfies the constraints coming from the
equations of motion and the heavy-quark symmetry\cite{heavyquark}.
They find that the ``Wandzura-Wilczek-type'' contribution\cite{ww}
(WW approximation), which corresponds to the valence quark
distribution, can be determined uniquely in analytic form in terms
of the ``effective mass'' ($\bar{\Lambda}$) of the meson state,
which is defined in the Heavy Quark Effective Theory
(HQET)\cite{hqet}. However, in Ref.\cite{qiao}, two extra
constraints for the B-meson wavefunction come from the heavy quark
limit, $m_b\to\infty$. Since the mass of b-quark is limited, such
condition might be too strong, and we will not take such limit in
our present calculation. In the following, we shall solve the two
equations that are derived from the equation of motion of the
light spectator quark on the basis of some physical
considerations.

\section{Equations under the WW approximation}

In HQET\cite{hqet}, the wavefunctions $\tilde{\Psi}_{\pm}(t,z^2)$
of the B meson can be defined in terms of the vacuum-to-meson
matrix element of the nonlocal operators:
\begin{equation}\label{hqeteq}
\langle 0 | \bar{q}(z) \Gamma h_{v}(0) |\bar{B}(p) \rangle = -
\frac{i f_{B} M}{2} {\rm Tr} \Bigg[ \gamma_{5}\Gamma \frac{1 +
\slash\!\!\! v}{2} \!\!\!\times \Bigg\{ \tilde{\Psi}_{+}(t,z^2)-
\slash\!\!\! z \frac{\tilde{\Psi}_{+}(t,z^2)
 -\tilde{\Psi}_{-}(t,z^2)}{2t}\Bigg\} \Bigg].
\end{equation}
Here, $z^{\mu}=(0, z^{-}, \mathbf{z}_\perp)$, $z^{2}= -
\mathbf{z}_{\perp}^{2}$, $v^{2} = 1$, $t=v\cdot z$, and $p^{\mu} =
Mv^{\mu}$ is the 4-momentum of the B meson with mass $M$.
$h_{v}(x)$ denotes the effective $b$-quark field. $\Gamma$ is a
generic Dirac matrix. The path-ordered gauge factors are implied
in between the constituent fields. Note that in the above
definition, the separation between the quark and the antiquark is
not restricted on the LC ($z^{2}=0$).

The effective mass $(\bar{\Lambda})$ is much smaller than the B
meson mass, so the light spectator quark in the B meson can be
treated as on mass shell with high precision. Based on the QCD
equation of motion for the light spectator quark, we can obtain a
set of equations for $\tilde\Psi_{\pm}(t,z^2)$ under the WW
approximation, i.e.
\begin{equation}\label{original1}
\frac{\partial \tilde\Psi_{-}(t,z^2)}{\partial t}-
\frac{\tilde\Psi_+(t,z^2)-\tilde\Psi_-(t,z^2)}{t}-\frac{z^2}{t}
\frac{\partial}{\partial z^2}[\tilde\Psi_+(t,z^2)
-\tilde\Psi_-(t,z^2)]= 0\;,
\end{equation}
and
\begin{equation}
\label{original2}\frac{\partial\tilde\Psi_+(t,z^2)}{\partial
t}-\frac{\partial\tilde\Psi_-(t,z^2)}{\partial
t}-\frac{\tilde\Psi_+(t,z^2)-\tilde\Psi_-(t,z^2)}{t}
+4t\frac{\partial\tilde\Psi_{+}(\omega,z^2)}{\partial z^2}= 0\;.
\end{equation}
When taking the LC limit $z^2\to 0$, the above two equations agree
well with the ones in Refs.\cite{beneke,descotes}. By doing the
Fourier transformation, $ \tilde{\Psi}_{\pm}(t, z^{2}) = \int
d\omega \ e^{-i \omega t} \Psi_{\pm}(\omega, z^{2})$,
Eqs.(\ref{original1},\ref{original2}) become that
\begin{equation}\label{eq:1}
\omega \frac{\partial \Psi_{-}(\omega,z^2)}{\partial \omega} + z^2
\left(\frac{\partial \Psi_{+}(\omega,z^2)}{\partial z^2}
-\frac{\partial \Psi_{-}(\omega,z^2)}{\partial
z^2}\right)+\Psi_{+}(\omega,z^2) = 0\;,
\end{equation}
and
\begin{equation}
\label{eq:2} \left(\omega\frac{\partial}{\partial \omega}+
2\right)[\Psi_{+}(\omega,z^2)-\Psi_{-}(\omega,z^2)]
+4\frac{\partial^{3}\Psi_{+}(\omega,z^2)}{\partial
\omega^{2}\partial z^2} = 0\;.
\end{equation}
Here, $\omega v^{+}$ has the meaning of the LC projection $k^{+}$
of the light-antiquark momentum in the B meson. The exact solution
of  Eqs.(\ref{eq:1},\ref{eq:2}) will give strong constraints on
the longitudinal and transverse momentum dependence of the B
wavefunction under the WW approximation.

\section{Solutions for the B wavefunction}

At present, one can not obtain the exact solution for the B-meson
wavefunctions $\Psi_{\pm}(\omega,z^2)$ only with
Eqs.(\ref{eq:1},\ref{eq:2}). Some prescriptions must be made.
Under the prescription that $\Psi_{\pm}(\omega,z^2)$ have the same
$z^2$-dependence and by taking the heavy quark limit
$m_b\to\infty$, Ref.\cite{qiao} obtained an analytic solution for
$\Psi_{\pm}(\omega,z^2)$. In the present paper, we do not take
such limit and try to solve Eqs.(\ref{eq:1},\ref{eq:2}) by taking
two other prescriptions.

Before solving Eqs.(\ref{eq:1},\ref{eq:2}), we define two
functions $\phi_\pm(\omega)$ in the following way,
\begin{eqnarray}\label{definephi}
\phi_\pm(\omega)&\equiv& \lim_{z^2\to 0}\Psi_{\pm}(\omega,z^2)\\
&=&\int d^2\mathbf{k}_\perp
\exp(i\mathbf{k}_\perp\cdot\mathbf{z}_\perp)\tilde
\Psi_\pm(\omega,\mathbf{k}_\perp^2)|_{\mathbf{z}_\perp\to 0}=\int
d^2\mathbf{k}_\perp\tilde\Psi_{\pm}(\omega,\mathbf{k}_\perp),\nonumber
\end{eqnarray}
where the second line shows that $\phi_\pm(\omega)$ are precisely
the B meson DAs. Taking the LC limit ($z^2\to 0$) in
Eq.(\ref{eq:1}), one can directly obtain a relation between the
two DAs $\phi_+(\omega)$ and $\phi_-(\omega)$ of the B meson,
\begin{equation}\label{relationap}
\omega \frac{\partial \phi_{-}(\omega)}{\partial \omega}
+\phi_{+}(\omega)=0,
\end{equation}
which agrees with the one obtained in Refs.\cite{beneke,descotes}.

Furthermore, one can prove that if the B-meson wavefunctions
$\Psi_{\pm}(\omega,z^2)$ can be constructed in the following way,
\begin{equation}\label{generalwf}
\Psi_{\pm}(\omega,z^2)=\phi_{\pm}(\omega)\cdot \chi(\omega,z^2)\ ,
\end{equation}
then $\chi(\omega,z^2)$ must be a function of the correlated
variable $[z^2\omega(\omega_0-\omega)]$\footnote{The special case
that $\chi(\omega,z^2)$ has nothing to do with the variable
$\omega$ has been discussed in Ref.\cite{descotes} and will not be
discussed here.}, where $\omega_0$ is the maximum value of
$\omega$. The ansatz (\ref{generalwf}) shows that the two
wavefunctions $\Psi_{\pm}(\omega,z^2)$ have the same
$z^2$-dependence $\chi(\omega,z^2)$. A boundary condition for
$\chi(\omega,z^2)$ can be derived from
Eqs.(\ref{definephi},\ref{generalwf}), i.e. $\lim_{z^2\to
0}\chi(\omega,z^2)=1$. From the above ansatz, we can solve
Eqs.(\ref{eq:1},\ref{eq:2}) exactly without taking the heavy quark
limit $m_b\to\infty$.

Substituting Eq.(\ref{generalwf}) into Eq.(\ref{eq:1}), we obtain
\begin{equation}\label{chixi}
\chi(\omega,z^2)=\xi(\omega\cdot\phi_{-}(\omega)\cdot
z^2)\equiv\xi(y),
\end{equation}
where $\xi(y)$ is the function of a single variable
$y=[\omega\cdot\phi_-(\omega)\cdot z^2]$. The explicit form of
$\phi_{-}(\omega)$ can be derived from
Eqs.(\ref{eq:2},\ref{chixi}), which reads
\begin{equation}\label{phizero}
\phi_{-}(\omega)=a_1\omega+a_0,\;\;a_1=-\frac{\xi(0)}{4\xi'(0)}
\end{equation}
where the parameter $a_0$ is to be determined. With the help of
the above results, Eq.(\ref{eq:2}) changes to
\begin{equation}\label{simeq2}
\frac{y}{2}\left(\frac{(a_0+2a_1\omega)^2}{(a_0+a_1\omega)(a_0+3a_1\omega)}
\right)\frac{d[a_1\sigma(y)+(1+a_1)\xi(y)]}{dy}
+[a_1\sigma(y)+(1+a_1)\xi(y)]=0,
\end{equation}
with $\sigma(y)=4y\xi''(y)+4\xi'(y)-\xi(y)$. Eq.(\ref{simeq2})
should be satisfied for all values of $\omega$, and then we have
to set,
\begin{equation}\label{finaleq}
a_1\sigma(y)+(1+a_1)\xi(y)=0.
\end{equation}
The non-trivial solution of the above equation is
\begin{equation}
\xi(y)=J_0(\sqrt{y/a_1}),\;\;\;\;\;\; (a_1<0)
\end{equation}
where $J_0$ is the zero-{\it th} normal Bessel function. Here
$a_1$ must be less than zero, because if $a_1=0$, then it will
lead to $\phi_{+}(\omega)\equiv 0$; and if $a_1>0$, then one may
obtain a self-contradictory result from Eq.(\ref{finaleq}):
$a_1=-K_0(0)/(4K'_{0}(0))\equiv 0$.

After doing the Fourier transformation for the transverse part,
$\tilde\Psi_{\pm}(\omega,\mathbf{k}_\perp)=\int
d^2\mathbf{z}_{\perp} \exp(-i \mathbf{k}_\perp\cdot
\mathbf{z}_\perp) \Psi_{\pm}(\omega,z^2)/(2\pi)^2$, the final
solution for the B-meson wavefunctions can be written as,
\begin{equation}\label{solution3}
\tilde\Psi_{\pm}(\omega,\mathbf{k}_\perp)=\phi_{\pm}(\omega)
\frac{\delta\left(\mathbf{k}_\perp^2-\omega(\omega_0-\omega)\right)}{\pi},
\end{equation}
where $\omega_0=(-a_0/a_1)$ is the maximum value of $\omega$ and
the normalized DAs take the form,
\begin{equation}\label{solution2}
\phi_{+}(\omega)=\frac{2}{\omega_0^2}\theta(\omega_0-\omega)\omega
; \;\;\;
\phi_{-}(\omega)=\frac{2}{\omega_0^2}\theta(\omega_0-\omega)
(\omega_0-\omega).
\end{equation}
The $\theta$ function guarantees that the maximum value of
$\omega$ is $\omega_0$. One may find that if taking
$(\omega_0=2\bar{\Lambda})$, the above solution for the B-meson
wavefunction agrees with the result in Ref.\cite{qiao}.
Eq.(\ref{solution3}) shows that the dependence on transverse and
longitudinal momenta is strongly correlated through the
combination $\mathbf{k}_\perp^2/[\omega(\omega_0-\omega)]$.
Similar transverse momentum behavior has been discussed in
Ref.\cite{bhl} and has been obtained by using the dispersion
relations and the quark-hadron duality\cite{halperin}, where they
stated that the $k_T$ dependence of the wavefunction depends on
the off-shell energy of the valence quarks, i.e. $\sim
\mathbf{k}_\perp^2/x(1-x)$, where $x$ is the momentum fraction
carried by the valence quarks. Under such kind of
$k_T$-dependence, by doing the Fourier transformation, we find
that the correlation between $\omega$ and $z^2$ for the transverse
part combines in a way like $[z^2\omega(\omega_0-\omega)]$. Thus
the above results has been proved completely and
Eqs.(\ref{eq:1},\ref{eq:2}) impose a strong constraint, i.e.
$\chi(\omega,z^2)=\chi[\omega\cdot(\omega_0- \omega)\cdot z^2]$,
on the B-meson wavefunction.

However, one may ask a question whether the ansatz
(\ref{generalwf}) is too strong. In order to study the B-meson
wavefunction more deeply, we assume that the B-meson wavefunction
$\Psi_{\pm}(\omega,z^2)$ can be constructed in a more general way
like,
\begin{equation}\label{bhleq}
\Psi_{\pm}(\omega,z^2)=\Phi_{\pm}(\omega)\cdot
[\rho(z^2)+\kappa(\omega)]\cdot
\chi[\omega\cdot(\omega_0-\omega)\cdot z^2]\ .
\end{equation}
Note here $\Phi_\pm(\omega)$ has no exact meaning of the
distribution amplitude. Substituting the above equation into
Eq.(\ref{eq:1}), one may find that the non-trivial solutions can
only be obtained under the following conditions,
\begin{eqnarray}
\frac{\rho'(z^2)}{\rho(z^2)}z^2[\Phi_+(\omega)-\Phi_-(\omega)]
+[\Phi_+(\omega)+\omega\Phi'_-(\omega)] = 0,\\
\omega\Phi_-(\omega)\kappa'(\omega)+
(\Phi_+(\omega)+\omega\Phi'_-(\omega))\kappa(\omega)=0,\\
\Phi_+(\omega)-\frac{\omega}{\omega_0-\omega} \Phi_-(\omega)= 0.
\end{eqnarray}
It can be seen that these equations have the following solutions,
\begin{eqnarray}
\rho(z^2)&=&f_1 (z^2)^C,\;\kappa(\omega)=f_3\omega^{-C}(\omega_0 -\omega)^{-C}\\
\Phi_+(\omega)&=&f_2\omega^{1+C}(\omega_0 -\omega)^C,
\Phi_-(\omega)=f_2\omega^{C}(\omega_0 -\omega)^{1+C},
\end{eqnarray}
where $C$ is arbitrary and $f_i(i=1,2,3)$ are undetermined
parameters that have nothing to do with $z^2$ and $\omega$.
Substituting these solutions into Eq.(\ref{eq:2}) and doing the
variable transformation $z^2\to x/[\omega(\omega_0-\omega)]$, we
obtain
\begin{equation}\label{goodeq}
\left[\omega^2 g_1(x)+(8\omega_0\omega-2\omega_0^2)
g_2(x)\right]=0,
\end{equation}
with
\begin{eqnarray}
g_1(x)&=&x{f_3}(-3\chi(x)-2(-6+x)\chi'(x)+
 4x(7\chi''(x)+2x\chi^{(3)}(x)))+\nonumber\\
&&x^C{f_1}((4C^2+8C^3-3x-2Cx)\chi(x)+2x((6+4C(4+3C)-\nonumber\\
&& x)\chi'(x)
+2x((7+6C)\chi''(x)+2x\chi^{(3)}(x)))),\nonumber\\
g_2(x)&=&x{f_3}(2\chi(x)+(-8+x)\chi'(x)-
4x(4\chi''(x)+x\chi^{(3)}(x)))-\nonumber\\
&& x^C{f_1}((4C^2(1+C)-(2+C)x)\chi(x)+
x( ( 8 +4C( 5 + 3C ) -\nonumber\\
&& x )\chi '(x)+4x( ( 4 + 3C ) \chi ''(x)+
 x\chi ^{(3)}(x)))),\nonumber
\end{eqnarray}
where $\chi '(x)=\partial\chi(x)/\partial x$, $\chi
''(x)=\partial^2\chi(x)/\partial^2 x$ and $\chi
^{(3)}(x)=\partial^3\chi(x)/\partial^3 x$. The functions
$g_i(i=1,2,3)$ must be set to zero so as to ensure that
Eq.(\ref{goodeq}) always be tenable with the variation of
$\omega$. And then we obtain that $\chi(x)$ must satisfy the
following equation,
\begin{eqnarray}\label{woowaa}
4x^2[f_1x^{C}+f_3]\chi''(x)+[8Cf_1x^{1+C}
+4f_1x^{C+1}-\nonumber\\
4f_3x]\cdot\chi'(x)+[4C^2f_1x^{C}-f_1x^{C+1}+f_3x]\chi(x)=0.
\end{eqnarray}
The solution of Eq.(\ref{woowaa}) from the condition of $f_3+f_1
x^{C}=0$ should be excluded, because when setting $f_3+f_1
x^{C}=0$, one may find that it will lead to
$\Psi_{\pm}(\omega,z^2)\equiv 0$. Under the condition that
$f_3+f_1 x^{C}\neq 0$ and by doing the transformation, $\chi(x)\to
\xi(x)/(f_3+f_1 x^{C})$, we obtain
\begin{equation}\label{bessel0eq}
4x\xi''(x)+4\xi'(x)-\xi(x)=0.
\end{equation}
Eq.(\ref{bessel0eq}) can be changed into a zero-{\it th} normal
Bessel function by doing the variable transformation,
$x\to\sqrt{-x}$ $(x<0)$, and then we get
$\chi(x)=J_0(\sqrt{-x})/(f_3+f_1 x^{C})$.

With the help of the above results, we finally obtain the solution
for the B-meson wavefunctions under the ansatz Eq.(\ref{bhleq}),
i.e.
\begin{eqnarray}\label{phiplus}
\tilde\Psi_{+}(\omega,\mathbf{k}_\perp^2)&=& \frac{2\omega}
{\pi\omega^2_{0}}\theta(\omega_0-\omega)\delta\left(\mathbf{k}_\perp^2-
\omega(\omega_0-\omega)\right),\\
\label{phiminus}
\tilde\Psi_{-}(\omega,\mathbf{k}_\perp^2)&=&\frac{2(\omega_0-\omega)}
{\pi\omega^2_{0}}\theta(\omega_0-\omega)
\delta\left(\mathbf{k}_\perp^2 -\omega(\omega_0-\omega)\right),
\end{eqnarray}
where the normalization condition, $\int d\omega
d^2\mathbf{k}_\perp \tilde
\Psi_{\pm}(\omega,\mathbf{k}_\perp^2)=1$, has been adopted.

\section{Physical consequence and discussion}

We have solved the equations which are derived from the equation
of motion of the light spectator quark by using two different
ansatz, i.e. Eq.(\ref{generalwf}) and Eq.(\ref{bhleq}). Under the
condition $\omega_0=2\bar{\Lambda}$, the results (see
Eqs.(\ref{solution3},\ref{solution2}) and
Eqs.(\ref{phiplus},\ref{phiminus})) for these two different ansatz
agree with the one obtained directly by taking the heavy quark
limit $m_b\to \infty$\cite{qiao}, i.e. these three prescriptions
are equivalent. Eqs.(\ref{phiplus},\ref{phiminus}) show that the
dependence on the transverse and longitudinal momenta is strongly
correlated through a non-factorizable combination
$\mathbf{k}_{\perp}^{2}/[\omega(\omega_0-\omega)]$. As has been
pointed out in Refs.\cite{qiao}, such kind of transverse momentum
dependence for the B-meson wavefunction has a slow-dumping with
oscillatory behavior at large transverse distances,
$\Psi_{\pm}(\omega,z^2) \sim
\cos\left(|\mathbf{z}_{\perp}|\sqrt{\omega(\omega_0-\omega)}
-\pi/4\right) /\sqrt{|\mathbf{z}_\perp|}$. This behavior is quite
different from that of the models with a simple Gaussian
distribution, which has a strong dumping behavior at large
transverse distances.

\begin{figure}
\centering
\begin{minipage}[c]{0.48\textwidth}
\centering
\includegraphics[width=2.9in]{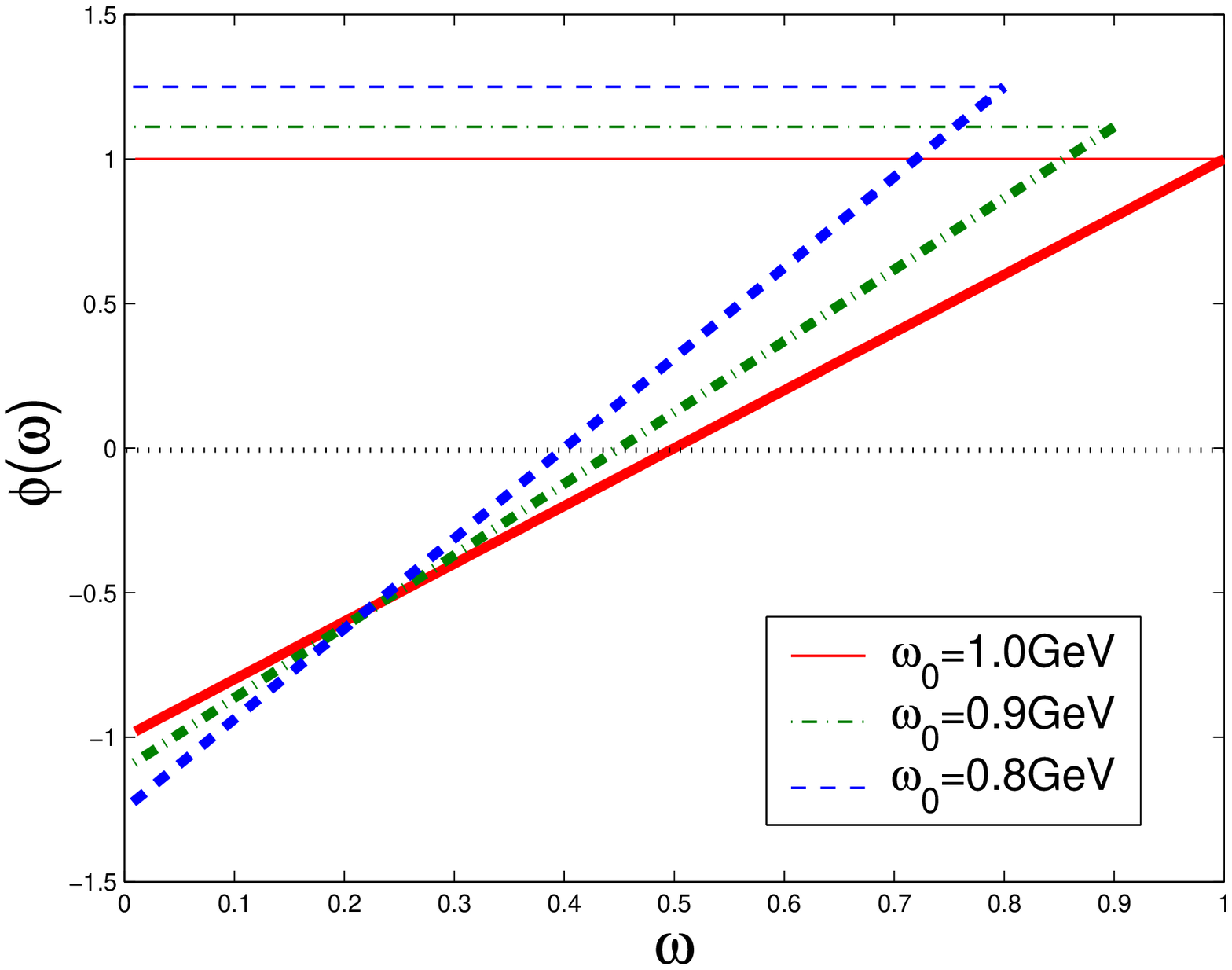}
(a)
\end{minipage}%
\begin{minipage}[c]{0.48\textwidth}
\centering
\includegraphics[width=2.9in]{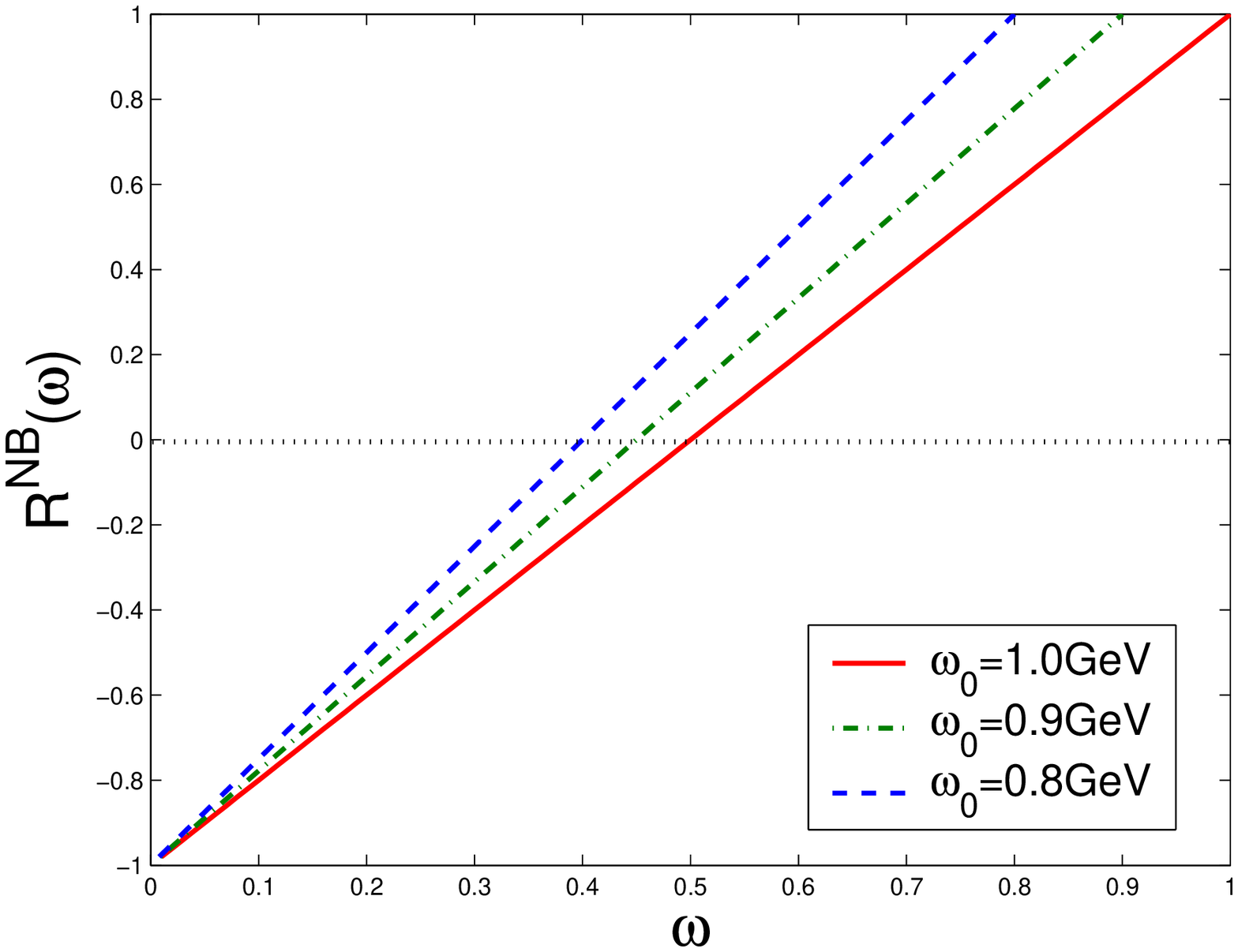}
(b)
\end{minipage}
\caption{The left diagram is the distributions of
$\phi^{NB}_B(\omega)$ and $\bar\phi^{NB}_B(\omega)$; the right
diagram is the distribution of the ratio
$R^{NB}(\omega)=\bar\phi^{NB}_B(\omega)/\phi^{NB}_B(\omega)$. In
the left diagram, the thicker same type lines are for
$\bar\phi^{NB}_B(\omega)$ and the thinner ones are for
$\phi^{NB}_B(\omega)$, respectively. } \label{phinb}
\end{figure}

Next, we make a discussion on the behavior of the obtained B-meson
distribution amplitudes. For latter convenience, we label the
B-meson distribution amplitudes in Eq.(\ref{solution2}) as
$\phi_\pm^{NB}(\omega)$. The value of $\omega_0$ in
Eq.(\ref{solution2}) can be taken as the continue threshold of the
light quarks in the B meson\cite{braun,continuum}, which is
usually taken to be in the interval $(0.8-1.0)$GeV\footnote{
observing that $\bar\Lambda\in(0.4,0.5)GeV$, $\omega_0$ roughly
agrees with the relation $\omega_0=2\bar\Lambda$.}. Other than
taking $\phi_\pm(\omega)$ directly into calculations, one usually
takes the combined form of $\phi_\pm(\omega)$, i.e.
\begin{equation}\label{definephib}
\phi_B(\omega)=\frac{\phi_{+}(\omega)+\phi_{-}(\omega)}{2} ,\;\;
\bar\phi_B(\omega)=\frac{\phi_{+}(\omega)-\phi_{-}(\omega)}{2}.
\end{equation}
Another typical definition for $\phi_B(\omega)$ and
$\bar\phi_B(\omega)$ can be found in Ref.\cite{lucai}, however one
may find that the qualitative conclusions are similar. We show the
distributions of $\phi^{NB}_B(\omega)$ and
$\bar\phi^{NB}_B(\omega)$ with varying $\omega_0$ in
Fig.(\ref{phinb}a) and the distributions of the ratio
$R^{NB}(\omega)=\bar\phi^{NB}_B(\omega)/\phi^{NB}_B(\omega)$ in
Fig.(\ref{phinb}b). One may observe that the value of
$\phi^{NB}_B(\omega)$ is always bigger than
$\bar\phi^{NB}_B(\omega)$ and the value of
$\bar\phi^{NB}_B(\omega)$ is negative in small $\omega$ regions.

\begin{figure}
\centering
\includegraphics[width=0.47\textwidth]{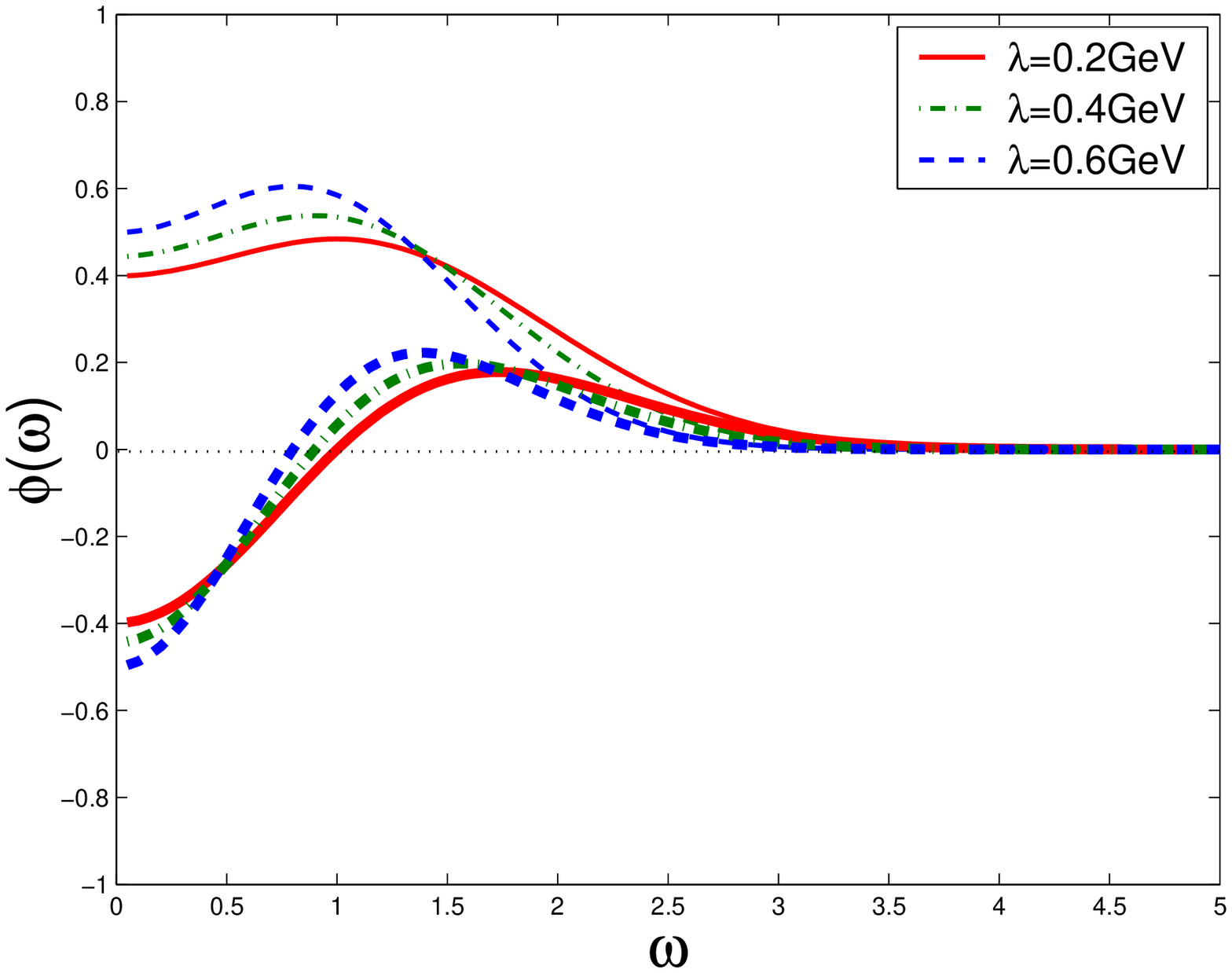}%
\hspace{0.4cm}
\includegraphics[width=0.47\textwidth]{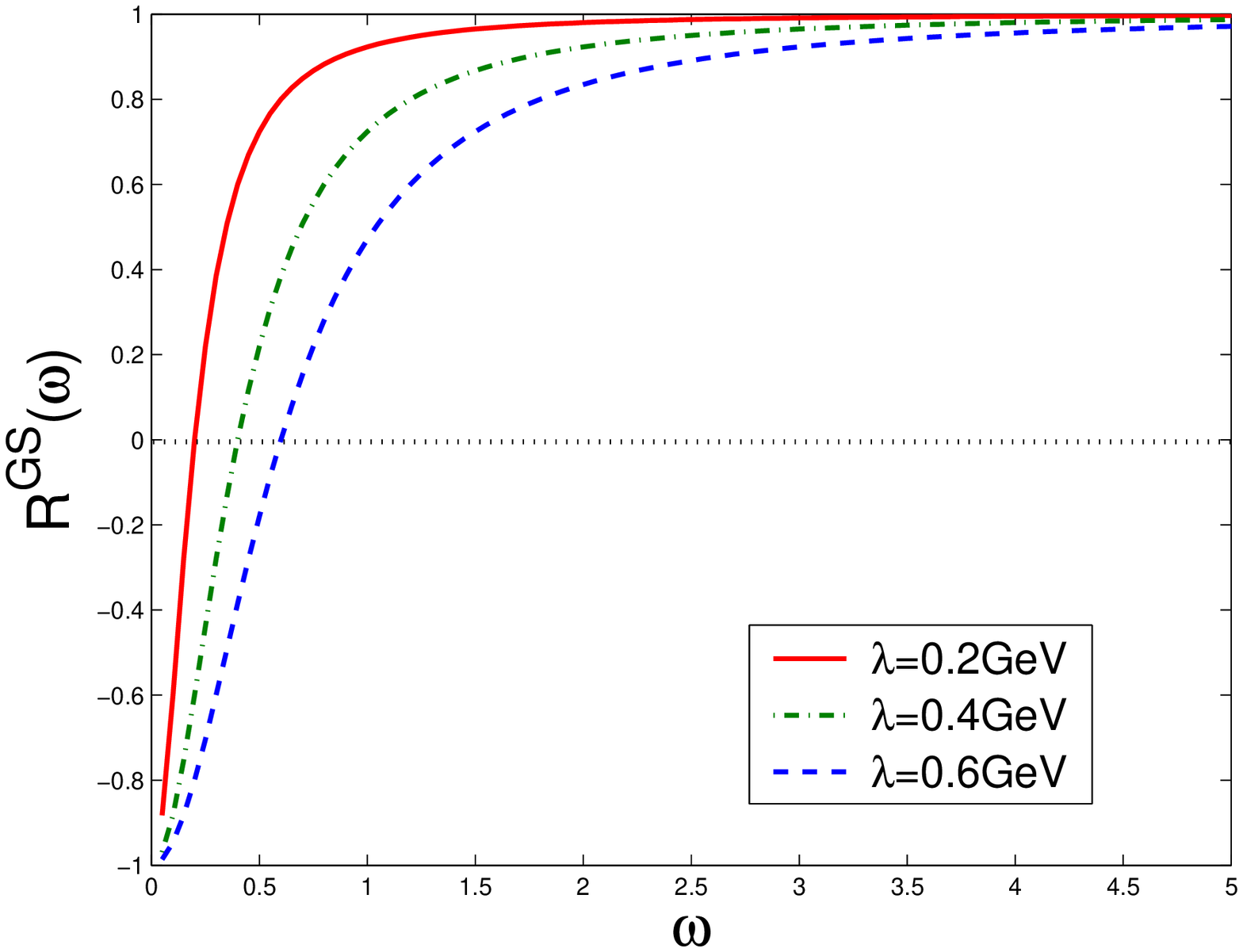}
\caption{The left diagram is the distributions of
$\phi^{GS}_B(\omega)$ and $\bar\phi^{GS}_B(\omega)$; the right
diagram is the distribution of the ratio
$R^{GS}(\omega)=\bar\phi^{GS}_B(\omega)/\phi^{GS}_B(\omega)$. In
the left diagram, the thicker same type lines are for
$\bar\phi^{GS}_B(\omega)$ and the thinner ones are for
$\phi^{GS}_B(\omega)$, respectively. } \label{phigs}
\end{figure}

Some other models for the B meson DAs have also been adopted in
the literature. In Ref.\cite{descotes}, the authors adopted the
model,
\begin{equation}\label{gd}
\phi^{GS}_{+}(\omega)=\sqrt{\frac{2}{\pi\lambda^2}}
\frac{\omega^2}{\lambda^2} \exp
\left(-\frac{\omega^2}{2\lambda^2}\right),\;
\phi^{GS}_{-}(\omega)=\sqrt{\frac{2}{\pi\lambda^2}}
\exp\left(-\frac{\omega^2}{2\lambda^2}\right),
\end{equation}
which are obtained under the prescription that the transverse part
and the longitudinal part of the B wavefunction can be strictly
factorized\cite{beneke,descotes}. In fact,
$\phi^{GS}_{\pm}(\omega)$ can be derived from
Eqs.(\ref{eq:1},\ref{eq:2}) by requiring $\partial
\chi(\omega,z^2)/\partial\omega=0$ together with the ansatz
Eq.(\ref{generalwf}). The parameter $\lambda$ can be determined
through $\lambda^2=4\partial\chi(\omega,z^2)/\partial z^2|_{z^2\to
0}$, whose value is of order $\Lambda_{QCD}$\cite{descotes}.
Inspired by a QCD sum rule analysis, Grozin and Nuebert
\cite{nuebert} have proposed a simple model for the distribution
amplitudes,
\begin{equation}\label{gn}
\phi^{GN}_+(\omega)=\frac{\omega}{\Omega^2_0}\exp\left
(-\frac{\omega}{\Omega_0}\right),\;
\phi^{GN}_-(\omega)=\frac{1}{\Omega_0}\exp
\left(-\frac{\omega}{\Omega_0}\right),
\end{equation}
where $\Omega_0=2\bar{\Lambda}/3$. Both $\phi^{GS}_\pm(\omega)$
and $\phi^{GN}_\pm(\omega)$ satisfy the relation
Eq.(\ref{relationap}). It means that $\phi^{NB}_\pm(\omega)$,
$\phi^{GS}_\pm(\omega)$ and $\phi^{GN}_\pm(\omega)$ all are
solutions of Eqs.(\ref{eq:1},\ref{eq:2}) as $z^2\to 0$.
$\phi^{NB}_\pm(\omega)$ and $\phi^{GN}_\pm(\omega)$ have different
behaviors, however they have the same asymptotic behavior that is
favored by most of the
calculations\cite{nuebert,cz,bdistribution,braun},
$\phi^{NB,GN}_{+}(\omega)\sim\omega$,
$\phi^{NB,GN}_{-}(\omega)\sim const$, as $\omega\to 0$.
$\phi^{GN}_\pm(\omega)$ and $\phi^{GS}_\pm(\omega)$ have similar
behavior (exponential form), but the asymptotic behavior for
$\phi_+$ is different, i.e. $\phi^{GS}_{+}(\omega)\sim\omega^2$,
as $\omega\to 0$. More clearly, we show the distributions of
$\phi^{GS}_B(\omega)$ and $\bar\phi^{GS}_B(\omega)$, and the ratio
$R^{GS}(\omega)=\bar\phi^{GS}_B(\omega)/\phi^{GS}_B(\omega)$ with
varying $\lambda$ in Fig.(\ref{phigs}). From Fig.(\ref{phigs}),
one may observe that similar to the case of $\phi^{NB}_B(\omega)$
and $\bar\phi^{NB}_B(\omega)$, $\bar\phi^{GS}_B(\omega)$ is also
comparable to $\phi^{GS}_B(\omega)$ in the endpoint regions.

One may observe from Figs.(\ref{phinb},\ref{phigs}) that the value
of $\phi^{NB,GS}_B(\omega)$ is always bigger than
$\bar\phi^{NB,GS}_B(\omega)$ and the value of
$\bar\phi^{NB,GS}_B(\omega)$ is negative in small $\omega$
regions, such behavior might lead to the total net contribution
from $\bar\phi^{NB,GS}_B(\omega)$ be much smaller than that of
$\phi^{NB,GS}_B(\omega)$. In literature, many authors (see
Refs.\cite{huangl,lihn,lihn1}) did the phenomenological analysis
with a single distribution amplitude $\phi_B$, setting
$\bar\phi_B=0$ (or strictly speaking, ignoring the contributions
from $\bar\phi_B$). However, since $\phi^{NB,GS}_+(\omega)$ and
$\phi^{NB,GS}_-(\omega)$ have a quite different endpoint behavior,
such difference maybe strongly enhanced by the hard scattering
kernel. The results in Refs.\cite{weiy,huangwu} for $B\to\pi$
transition form factor confirm this observation. Especially in
Ref.\cite{huangwu}, by comparing the PQCD results in the large
recoil regions with those obtained from the QCD light-cone sum
rules and the extrapolated lattice QCD, the authors give a
detailed analysis on the consistent calculation of the $B\to\pi$
transition form factor in the whole physical region. Their results
show that a better slope of the PQCD results can be obtained by
taking both $\phi_B$ and $\bar\phi_B$ into consideration. In fact,
a discussion on using a single $\phi_B$ is given in
Ref.\cite{lihn}, which is based on the assumption that $\phi_-$
vanishes at the both ends of the momentum $\omega$. There is
therefore no convincing motivation for setting $\bar\phi_B=0$ and
such an approximation may lead to unreliable results. Since for
the endpoint region, the contribution from the hard scattering
part might be big, which is the case of $B\to\pi$ and
$B\to\rho$\cite{mahajan} transition form factors, so we argue that
both $\phi_B$ and $\bar\phi_B$ should be kept for a better
understanding of the B physics.

\section{Conclusion}

In summary, we have solved the B meson wavefunction based on the
equations derived from the vacuum-to-meson matrix element of the
nonlocal operators in the heavy quark effective theory and from
the equation of motion of the light spectator quark in the B
meson. Our analysis shows that the equation of motion of the light
spectator quark can impose a strong constraint on the B-meson
wavefunctions $\Psi_\pm(\omega,z^2)$. For example, the function
$\chi(\omega,z^2)$ depends only on a single combined variable
$[\omega\cdot(\omega_0-\omega)\cdot z^2]$, if assuming
$\Psi_\pm(\omega,z^2)=\phi_\pm(\omega)\chi(\omega,z^2)$. According
to our discussion, the distribution amplitude $\bar\phi_B(\omega)$
of the B meson can not be safely neglected for a better
understanding of the B decays.

\begin{center}
\section*{Acknowledgements}
\end{center}

This work was supported in part by the Natural Science Foundation
of China (NSFC).\\


\begin{thebibliography}{99}

\bibitem{collinear} G.P. Lepage and S.J. Brodsky, Phys.Lett.
B{\bf 87}, 359(1979); Phys.Rev. D{\bf 22}, 2157(1980); A.V.
Efremov and A.V. Radyushkin, Phys.Lett. B{\bf 94}, 245(1980).

\bibitem{bdistribution} B.O. Lange and M. Neubert, Phys.Rev.Lett.
{\bf 91}, 102001(2003).

\bibitem{braun} V.M. Braun, D.Y. Ivanov and G.P. Korchemsky, Phys.Rev. D{\bf
69}, 034014(2004).

\bibitem{lange} B.O. Lange, Eur.Phys.J. C{\bf 33}, S259(2004).

\bibitem{qiao0} H. Kawamura, J. Kodaira, C.F. Qiao and K. Tanaka,
Phys.Lett. B{\bf 523}, (111)2001, Erratum-ibid. B{\bf 536}, 344
(2002).

\bibitem{beneke} M. Beneke, T. Feldmann, Nucl.Phys. B{\bf 592},
3(2001).

\bibitem{descotes} S.D. Genon and C.T. Sachrajda, Nucl.Phys. B{\bf
625},239(2002).

\bibitem{libwave} H.N. Li and H.S. Liao, Phys.Rev. D{\bf 70},
074030(2004).

\bibitem{kt} J. Botts and G. Sterman, Nucl.Phys. B{\bf 225},
62(1989); H.N. Li and G. Sterman, Nucl.Phys. B{\bf 381},
129(1992); M. Nagashima and H.N. Li, Phys.Rev. D{\bf 67},
034001(2003), and references therein.

\bibitem{endpoint} A. Szczepaniak, E.M. Henley and S. Brodsky,
Phys.Lett. B{\bf 243}, 287(1990).

\bibitem{qiao} H. Kawamura, J. Kodaira, C.F. Qiao and K. Tanaka,
Nucl.Phys. B(Proc.Suppl.){\bf 116}, 269(2003); Mod.Phys.Lett.
A{\bf 18}, 799(2003).

\bibitem{heavyquark} N. Isgur and M.B. Wise Phys. Lett. B{\bf 232}, 113
(1989); N. Isgur and M.B. Wise Phys. Lett. B{\bf 237}, 527(1990);
E. Eichten and B. Hill, Phys.Lett. B{\bf 234}, 511(1990).

\bibitem{ww} S. Wandzura and F. Wilczek, Phys. Lett. B{\bf 72}, 195(1977).

\bibitem{hqet} H. Georgi, Phys.Lett. B{\bf 240}, 447(1990); A.F. Falk, H. Georgi,
B. Grinstein and M.B. Wise, Nucl.Phys. B{\bf 343},1(1990); M.
Neubert, Phys. Rept. {\bf 245}, 259 (1994).

\bibitem{bhl} T. Huang, {\it in Proceedings of XXth International
Conference on High Energy Physics}, Madison, Wisconsin, 1980,
edited by L.Durand and L.G. Pondrom, AIP Conf.Proc.No. 69(AIP, New
York, 1981), p1000; S.J. Brodsky, T. Huang and G.P. Lepage, in
{\it Particles and Fields}, Vol.2, Proceedings of the Banff Summer
Institute, Banff, Alberta, 1981, edited by A.Z. Capri and A.N.
Kamal (Plenum, New York, 1983), P143.

\bibitem{halperin} I.E. Halperin, A. Zhitnitsky, Phys.Rev. D{\bf 56},
184(1997).

\bibitem{nuebert} A.G. Grozin and M. Neubert, Phys.Rev. D{\bf 55},
272(1997).

\bibitem{cz} V.L. Chernyak and A.R. Zhitnitsky, Phys.Rep. {\bf
112}, 173(1984).

\bibitem{continuum} P. Ball and V.M. Braun, Phys.Rev. D{\bf 49},
2472(1994); M. Neubert, Phys.Rev. D{\bf 45}, 2451(1992); E. Bagen,
P. Ball, V.M. Braun and H.G. Dosch, Phys.Lett. B{\bf 278},
457(1992); M. Neubert, Phys.Rep.{\bf 245}, 259(1994).

\bibitem{lucai} C.D. Lu and M.Z. Yang, Eur.Phys.J. C{\bf 28}, 515(2003).

\bibitem{lihn1} H.N. Li, Phys.Rev. D{\bf 52}, 3958(1995); H.N. Li
and B. Melic, Eur.Phys.J. C{\bf 11}, 695(1999); C.D. L\"u, K. Ukai
and M. Yang, Phys.Rev. D{\bf 63}, 074009(2001); M. Dahm, R. Jaco
and P. Kroll, Z.Phys. C{\bf 68}, 595(1995).

\bibitem{huangl} T. Huang and C.W. Luo, Commun.Theor.Phys. {\bf 22},
473(1994); A. Szczepaniak, E.M. Henley and S.J. Brodsky,
Phys.Lett. B{\bf 243}, 287(1990); S.J. Brodsky, SLAC-PUB-5529;
S.J. Brodsky, SLAC-PUB-5917.

\bibitem{lihn} Y.Y. Keum, H.N. Li and A.I. Sanda,
Phys.Rev. D{\bf 63}, 054008(2001).

\bibitem{weiy} Z.T. Wei and M.Z. Yang, Nucl.Phys. B{\bf 642},
263(2002).

\bibitem{huangwu} T. Huang and X.G. Wu, hep-ph/0412417.

\bibitem{mahajan} N. Mahajan, hep-ph/0405161.

\end{thebibliography}
\end{document}